\documentstyle[12pt]{article}
\def\ddt{\frac{\partial}{\partial t}}
\def\nup{{\nu^\prime}}
\def\p{{\mbox{\boldmath$p$}}}
\def\r{{\mbox{\boldmath$r$}}}
\def\nm{{\nu\mu}}
\def\mn{{\mu\nu}}
\def\mup{{\mu^\prime}}
\def\om{\omega}
\def\chipp{\chi^{\prime\prime}}
\def\k{{\mbox{\boldmath$k$}}}
\def\g{{\bf g}}
\def\G{{\mbox{\boldmath$G$}}}
\def\kp{{\k^\prime}}
\def\dern0{\frac{\partial n^0}{\partial\varepsilon_p}}
\def\pp{{\p^\prime}}
\def\rp{{\r^\prime}}
\def\ppp{\p^{\prime\prime}}
\def\rpp{\r^{\prime\prime}}
\def\v{{\rm v}}

\def\eps{\varepsilon}

\author{D.~Kiderlen\\Physik-Department, TU M\"unchen, D-85747 Garching\\
        and\\
        National Superconducting Cyclotron Laboratory\\
        Michigan State University, East Lansing, Michigan 48824
        \vspace{0.8cm}}
\title{Dynamical fluctuations in the one particle density
-- comparison of different approaches\vspace{0.5cm}}
\date{March 1, 1995}
\begin{document}

\maketitle

\begin{abstract}
\noindent
Diffusion coefficients are obtained from linear response functions and from the
quantal fluctuation dissipation theorem. They are compared with the results of
both the theory of hydrodynamic fluctuations by Landau and Lifschitz as well as
the Boltzmann-Langevin theory. Sum rules related to conservation laws for total
particle number, momentum and energy are demonstrated to hold true for
fluctuations and diffusion coefficients in the quantum case.
\end{abstract}

\section{Introduction}

For a theoretical description of phenomena like nuclear multifragmentation mean
field theories have to be extended to include the dynamics of fluctuations. In
the case of the semi-classical Boltzmann-\"Uhling-Uhlenbeck (BUU) equation a
first step was made by Bauer et al.~\cite{BBdG}. Their numerical realization of
the collision term corresponds to adding to the BUU collision term a further
contribution. The average of the latter should vanish since BUU is meant to
describe the effect of the collisions on average. In this sense the additional
term acts like a fluctuating force. However, in \cite{BBdG} detailed properties
of the former as its second moment (or correlation function) were not
investigated. Moreover, in a later work \cite{CBCR} the fluctuations that
are obtained within the method of \cite{BBdG} were judged unsatisfactory.

Possibilities to extend theories for average dynamics were proposed before
by Landau and Lifschitz \cite{LL6} for hydrodynamics,
by Abrikosov and Khalatnikov \cite{AbKh} for Landau Fermi Liquid theory and
by Bixon and Zwanzig \cite{BZ} for the Boltzmann equation.
These authors augment the equations for the averages with a fluctuating term
and derive explicitly its correlation function, albeit in different ways.

The task to describe heavy ion collisions, especially multifragmentation,
initiated the more recent approaches by Ayik and Gr\'egoire \cite{AG}, Randrup
and Remaud \cite{RR} and Hofmann, the author and Tsekhmistrenko \cite{HKT}.
In \cite{RR} and \cite{HKT} the equation of motion for the fluctuations (the
second cumulants) of the distribution function (the Wigner transform of the one
particle density matrix) is addressed, as first step.
The inhomogeneity of this equation, called diffusion coefficient in analogy to
Fokker Planck equations, can be related to the correlation function of the
fluctuating force. The latter is derived in \cite{AG} by investigating the
hierarchy of equations for the n-particle densities. It is found to coincide
with the one obtained from the expression of \cite{RR} for the diffusion
coefficient -- for the case that the equation for the average distribution
function is given by the BUU equation with a Markovian collision term.

In the approach of \cite{HKT} the diffusion coefficient is determined to
guarantee that stationary solutions of the equations of motion for the
dynamical fluctuations are, for a stable system, equal to the equilibrium
values. The latter can be determined by means of the fluctuation dissipation
theorem \cite{CalWel} to include quantum effects beyond Pauli blocking. In a
recent work \cite{KH} these quantum effects were investigated in the density
range between one tenth of up to two times the saturation density. They were
found to be sizable for temperatures below 15 MeV (This temperature range is
of special interest as it is the one in which, in a certain density range
below saturation density, homogeneous nuclear matter can become mechanically
unstable against isothermal density variations, in other words, the temperature
range of the isothermal spinodal region of nuclear matter.).

In the approaches of \cite{AG} and \cite{RR} -- both now known under the name
Boltzmann-Langevin theory -- correlation function of the fluctuating force and
diffusion coefficient, respectively, are determined as functionals of the
distribution function. The latter is allowed to be the one of a non
equilibrium state of the system. However, the numerical evaluation of these
functionals turns out to be too time consuming as to presently allow for an
application to a realistic (six dimensional) system. Therefore, recently
\cite{RanAy} the expression for the diffusion coefficient was proposed to be
evaluated at the equilibrium distribution, similar to the approach of
\cite{HKT}.

It is the main aim of the present paper to work out the relation of the
approach of \cite{HKT} based on the (quantal) fluctuation dissipation theorem
to those of Landau--Lifschitz and Bixon--Zwanzig as well as to the
Boltzmann-Langevin theory. The connection to the latter will be established for
an equilibrium state in the case that the classical form of the fluctuation
dissipation theorem is applied.

In Boltzmann-Langevin theory sum rules reflecting conservation laws for total
particle number, momentum and energy are fulfilled by the expressions obtained
for the correlation function of the fluctuating force \cite{AG} and for the
diffusion coefficients \cite{BCR}. These sum rules will be demonstrated to
hold true also for fluctuations and diffusion coefficient found in the
approach of \cite{HKT} -- in the classical limit as well as for the quantum
corrections. A proper interpretation will be given for previous results
\cite{HKT} which seem to violate these sum rules.

\section{Review of general formalism}

In this section the basic ideas of the approach of \cite{HKT,KH} will be
reviewed using a more generally applicable formulation. It will be explained in
which way equations of motion for the fluctuations are obtained starting from
the equations for the first moments or averages. The former turn out to be
inhomogeneous. The knowledge of this inhomogeneity, the diffusion coefficient,
can serve as the first step in the determination of the correlation function of
the fluctuating force. Moreover, the equations obtained can be used to study
the behaviour of the dynamical fluctuations, at least in the range of small
amplitudes \cite{thesis}.

The starting point are the equations of motion for averages $A_\nu$ of a set of
dynamical variables $\hat A_\nu$,
\begin{equation}\label{avofop}
A_\nu (t)= \left<\hat A_\nu\right>_t
\; ,
\end{equation}
the system is to be described with. These equations are assumed to have the
form
\begin{equation}\label{eomav}
\ddt A_\nu + {\cal D}_{\nu\nup} A_\nup + c_\nu= 0
\; .
\end{equation}
Here and in the following summation over repeated indices is understood to be
performed. The linearity in $A_\nu$ may have been obtained by linearization.
The $c_\nu$ are independent of $A_\nu$ and symbolize terms containing either
external fields or zero order contributions remaining from the
linearization procedure or both.

To have a specific example in mind let us look at the case where the system is
described by the distribution function (which is the Wigner transform of the
one particle density matrix) $n_\p(\r,t)$. In the semi classical limit the
equation of motion of the latter is the BUU equation which is nonlinear in
$n_\p(\r,t)$. One, therefore, has to perform a linearization to obtain an
equation that can be written in the form of (\ref{eomav}). The averages
$A_\nu$ would then be the deviation $\delta n_\p(\r,t)$ of the distribution
function from the reference distribution chosen, e.g.~the one of global
equilibrium,
\begin{equation}\label{rela}
A_\nu (t)\leftrightarrow \delta n_\p (\r,t) = n_\p (\r,t)-n^0(\varepsilon^0_p)
\; .
\end{equation}
The index $\nu$ stands for position $\r$ and momentum $\p$ and eventually
for a spin-isospin index, the latter being suppressed throughout the paper.
So far, the reference distribution might be the one of a local equilibrium or
some other state as well. One simply would obtain expressions for
${\cal D}_{\nm}$ and $c_\nu$ different from those found when linearizing around
global equilibrium.

The fluctuations of the dynamical variables are defined as:
\begin{equation}\label{deffluct}
\sigma_\nm (t)=
\left<\frac{1}{2} \left[\hat A_\nu\, ,\,\hat A_\mu^\dagger\right]_+\right>_t -
A_\nu (t) A_\mu^\ast (t)
\end{equation}
{}From this form the property
\begin{equation}\label{hermsig}
\sigma_\nm = \sigma_\mn^\ast
\end{equation}
follows even for non hermitian $\hat A_\nu$, as one has to deal with
considering the spatial Fourier transform of $\delta n_\p(\r,t)$ .

Since the equations of motion (\ref{eomav}) for the averages $A_\nu$ have been
assumed linear, the corresponding ones for the $\hat A_\nu$ differ from the
former only by a term, called the fluctuating force, whose average vanishes.
Thus the equations for the fluctuations are found to read
\begin{equation}\label{eomfluct}
\ddt \sigma_\nm +  {\cal D}_{\nu\nup} \sigma_{\nup\mu} +
\Big({\cal D}_{\mu\mup} \sigma_{\mup\nu}\Big)^\ast = 2\, d_\nm
\; ,
\end{equation}
where the diffusion coefficient $d_\nm$ includes all terms arising from the
fluctuating force. In general, $d_\nm$ depends on the momentary state of the
system: looking at the hierarchy of n-particle densities mentioned before it
becomes clear that $d_\nm$ stands for terms containing the correlated part of
the three particle density. To calculate these terms exactly would be
equivalent to solving the full hierarchy which is, of course, not possible.
One, therefore, has to use approximations. The one made in the present
approach is the following. For a stable system $d_\nm$ is chosen in such a
way that the stationary solution of (\ref{eomfluct}) is equal to the
equilibrium fluctuations; for an unstable system $d_\nm$ is obtained \cite{KH}
as the analytic continuation of the expressions found for the stable
system. This is achieved by setting
\begin{equation}\label{defdifco}
2 d_\nm = {\cal D}_{\nu\nup} \sigma_{\nup\mu}^{\rm st} +
\Big({\cal D}_{\mu\mup} \sigma_{\mup\nu}^{\rm st}\Big)^\ast
\end{equation}
where, for the stable system, $\sigma^{\rm st}_\nm$ stands for the equilibrium
fluctuations, while it represents the analytical continuation of the latter in
the unstable case.

The diffusion coefficient obtained in this way depends on both the equilibrium
properties -- via $\sigma^{\rm st}_\nm$ -- and the dynamics of the system --
via ${\cal D}_\nm$. Let us look again at the example previously discussed where
the linearized BUU-equation is the one from which ${\cal D}_\nm$ is to be
obtained. In this case ${\cal D}_\nm$ containes information not only about
the drift terms (involving the equilibrium mean field) but also about the
collision term and thus about the differential scattering cross section.
Therefore, the diffusion coefficient depends on both the mean field and the
residual interaction causing the two particle collisions.

Clearly, this choice for the diffusion coefficient is an approximation to
the exact term appearing as inhomogeneity in (\ref{eomfluct}). However,
in the case that the system approaches equilibrium, the value of the latter
approaches the one of the former, which thus can serve as an approximation
valid for systems not too far from equilibrium.

Since ${\cal D}_\nm$ is determined by the form (\ref{eomav}) of the equation
of motion for $A_\nu$, what is left in order to calculate the diffusion
coefficient is the determination of the equilibrium fluctuations (and their
analytical continuation to the unstable case). For this purpose the relation
of the latter to dissipative processes in the average dynamics as expressed in
the fluctuation dissipation theorem is exploited. The latter provides the
connection between the equilibrium fluctuations and the corresponding response
function which can be obtained from the equation of motion for the averages
$A_\nu$ as follows.

Introducing an external field $U^{\rm ext}$ in such a way that the change in
the Hamiltonian due to $U^{\rm ext}$ is given by
\begin{equation}\label{hint}
\delta H = \hat A^\dagger_\nup\, U^{\rm ext}_\nup (t)
\; ,
\end{equation}
one defines the response function as the negative functional derivative of the
Fourier transform $\left<\hat A_\nu\right>(\om)$ of the average $A_\nu(t)$
with respect to the external field:
\begin{equation}\label{chimakro}
\chi_\nm (\om) =
-\left.\frac{\delta\left<\hat A_\nu\right>(\om)}{\delta U^{\rm ext}_\mu (\om)}
\right\vert_{U^{\rm ext}=0}
\end{equation}
This derivative can be calculated from the equation of motion of the $A_\nu$.
The dissipative processes are related to the quantity
\begin{equation}\label{defchippz}
\chipp_\nm (\om)=\frac{1}{2i}\Big(\chi_\nm (\om)-\chi_\mn (\om^\ast)^\ast\Big)
\end{equation}
which, therefore, is usually called, for real arguments $\om$, the dissipative
part of the response function (\ref{chimakro}).

According to the fluctuation dissipation theorem \cite{CalWel} the
$\sigma^{\rm st}_\nm$ are given by the sum of two contributions:
\begin{equation}\label{fdtgen}
\sigma^{\rm st}_\nm = \tilde\sigma^{\rm st}_\nm + \Delta\sigma_\nm
\; ,
\end{equation}
to be explained in the following.

The first contribution on the right hand side shall be given by
\begin{equation}\label{sigtilst}
\tilde\sigma^{\rm st}_\nm= \hbar\,
\int_C \frac{d\om}{2\pi} \coth\left(\frac{\hbar\om}{2T}\right)\chipp_\nm (\om)
\; ,
\end{equation}
where T stands for the temperature.
The integral in (\ref{sigtilst}) is to be taken along a contour $C$. For a
stable system $C$ lies on the real axis. In this case the stationary solution
$\sigma^{\rm st}$ (\ref{fdtgen}) can be interpreted as the equilibrium values
of the fluctuations and (\ref{fdtgen}) with (\ref{sigtilst}) is one version of
the fluctuation dissipation theorem. For an unstable system, $C$
has to be deformed in such a way that it lies above all poles of the response
function $\chi(\om)$ and below all poles of the function $\chi(\om^\ast)$ in
the complex frequency plane, crossing the line ${\rm Re}\,\om=0$ in the
interval $-2\pi\, T/\hbar <{\rm Im}\, \om< 2\pi\, T/\hbar$.

Let us turn now to the second contribution to $\sigma^{\rm st}$, i.e.~to
$\Delta\sigma_\nm$. Its physical origin lies in the possibility that, due to
conserved quantities, an initially equilibrated system does not relax to the
new thermal equilibrium (grand-, micro- or canonic) after an adiabatic switch
on of an external perturbation.

To shed some more light on this let us look at the case where the equilibrium
distribution is assumed to be the one for the canonical ensemble. Then
$\Delta\sigma_\nm$ is given \cite{brenig} by the positive semi definite
matrix\footnote{A matrix M is called positive semi definite, if
$\sum_{\nu,\mu} M_\nm x_\nu x_\mu \ge 0$ for any real $x_\nu$.}
\begin{equation}\label{dels}
\Delta\sigma_\nm= T \Big(\chi^{\rm T}_\nm - \chi_\nm(\om=0)\Big)
\end{equation}
where the isothermal susceptibility $\chi^{\rm T}$ can be calculated from the
behaviour of the equilibrium values of $A_\nu$ as functions of the
external field:
\begin{equation}\label{chiisotmakro}
\chi^{\rm T}_\nm = -\left.\left(\frac{\delta <\hat A_\nu >^{\rm eq}}
{\delta U^{\rm ext}_\mu}\right)_{\rm T}\right\vert_{U^{\rm ext}=0}
\end{equation}
Here, the change in the equilibrium values of $A_\nu$ has to be taken under the
subsidiary condition of a constant temperature. The difference between
isothermal suszeptibility and static response as appearing in (\ref{dels}) can
then be expressed in the following way: using the
isothermal suszeptibility as scalar product in the space of dynamical variables
one can determine an orthogonal basis $\{{\cal V}_c\}$ in the subspace of those
variables representing conserved quantities. One then finds \cite{kubo2}:
\begin{equation}\label{chitmchio}
\chi^{\rm T}_\nm - \chi_\nm(\om=0)=
\sum_c {\chi^{\rm T}_{\nu {\cal V}_c}\chi^{\rm T}_{{\cal V}_c\mu}\over
        \chi^{\rm T}_{{\cal V}_c\, {\cal V}_c }}
\end{equation}
where the sum on the right hand side extends over all elements ${\cal V}_c$
of the basis. This formula clearly states the relation of $\Delta\sigma$ to
the existence of conserved quantities.

These considerations are valid for a system in contact with a heat bath
preserving the temperature. However, if the system
to be investigated has no heat bath, one might conclude that, for such a case,
$\Delta\sigma_\nm$ has to be modified as compared to (\ref{dels}), like it was
done in \cite{KH}.

In the present paper the following point of view is taken: Quite generally a
contribution $\Delta\sigma$ to $\sigma^{\rm st}$ must be expected to exist;
there will be no calculation of the former; for the comparison with the other
approaches only $\tilde\sigma^{\rm st}$ will be used. It will be found that the
former are consistent with $\Delta\sigma=0$. Actually, we would like to use
this result as justification to identify $\tilde\sigma^{\rm st}$
(\ref{sigtilst}) with $\sigma^{\rm st}$  (\ref{fdtgen}). Nevertheless,
$\tilde\sigma^{\rm st}$ will be distinguished from $\sigma^{\rm st}$ by
keeping the tilde whenever the expression on the right hand side of
(\ref{sigtilst}) is to be referred to.

At this stage it is possible to briefly compare with the approach of Bixon and
Zwanzig \cite{BZ}. (For more details see the appendix A.) These authors address
a stable system with the Boltzmann equation as the one governing the average
dynamics of the distribution function. For comparison of the results of the two
methods one, therefore, has to define ${\cal D}_\nm$ and $c_\nu$ by writing the
linearized Boltzmann equation in the form (\ref{eomav}) with (\ref{rela}). One
finds the product of the expression on the right hand side of (\ref{defdifco})
with a $\delta-$function in time to equal the expression of \cite{BZ} for the
correlation function of the fluctuating force. This is consistent with the
assumption of a vanishing memory for the fluctuating force made in \cite{BZ}.
The value for the equilibrium fluctuations used in \cite{BZ} is the same as the
one obtained from (\ref{sigtilst}), provided the classical limit of the latter
is taken and the response function is calculated from the Boltzmann equation
(Please notice, that in the form of the latter used in \cite{BZ} no particle
interaction is taken into account.). This can be concluded from the results in
\cite{KH}.

\section{Fluctuations for Hydrodynamics}

In this section the method described in the previous one for the determination
of the diffusion coefficients is applied to hydrodynamics. Agreement with
the method of Landau and Lifschitz \cite{LL6} will be established. For the case
of hydrodynamics the index $\nu$ stands for a continuous variable and a
discrete one. The former is given by the spatial position $\r$ or --
in the corresponding Fourier space -- the wave vector $\k$. The discrete one is
meant to tell whether $A_\nu$ is the deviation $\delta\rho$ of the mass density
from the equilibrium value $\rho_0$, the deviation
$\delta q = T_0\rho_0\,\delta s$ of the density of heat or the momentum density
$\g$. (Here $s$ is the entropy per unit mass and $T_0$ is the equilibrium value
of the temperature.)

The linearized hydrodynamic equations for the dynamical variables, i.e.~those
corresponding to $\hat A_\nu$, read (see e.g.~\S 132 of \cite{LL6}):
\begin{equation}\label{linhyd1}
\ddt\,\delta\hat\rho\,(\k,t) + i\,\k\, \hat\g\,(\k,t) = 0
\end{equation}
\begin{equation}\label{linhyd2}
\ddt\, {\rm \hat g}_i\,(\k,t)+i\, k_j\, \Pi_{ij}\,(\k,t)=i\, k_j\, s_{ij}(\k,t)
\end{equation}
\begin{equation}\label{linhyd3}
\ddt\, \delta\hat q\,(\k,t) + i\, \k\,{\bf j}_T\, (\k,t) = -i\, \k\, \G(\k,t)
\end{equation}
Here beside the usual stress tensor
\begin{equation}\label{hydmod1}
\Pi_{ij}(\k,t)=\delta_{ij}\, P(\k,t) - i\Big(D_t\,(k_j\,{\rm\hat g}_i +
k_i\,{\rm\hat g}_j)+ (D_l- 2 D_t)\,\delta_{ij}\,\k\,\hat\g\Big)
\end{equation}
and heat current
\begin{equation}\label{hydmod2}
{\bf j}_T\,(\k,t) = -i\, \k\, \kappa\, \delta T(\k,t)
\; ,
\end{equation}
where $P$ is the pressure, $D_l$ and $D_t$ are the longitudinal and transversal
diffusion constants, respectively, and $\delta T$ is the deviation of the
temperature from $T_0$, the fluctuating forces $s_{ij}$ and $\G$ appear. In the
form suitable for later purposes their correlation functions are given
\cite{LL6} by:
\begin{equation}\label{corrs}
\left<{1\over 2}\Big[s_{ij}(\k,\om)\, ,\, s_{mn}(-\kp,t)\Big]_+\right>=
(2\pi)^3\delta(\k-\kp)\, 2 E(\om)\, e^{i\om t} \rho_0 D_{ijmn}
\end{equation}
\begin{equation}\label{corrg}
\left<{1\over 2}\Big[G_i(\k,\om)\, ,\, G_j(-\kp,t)\Big]_+\right>=
(2\pi)^3\delta(\k-\kp)\, 2 E(\om)\, e^{i\om t} \kappa\, T_0\, \delta_{ij}
\end{equation}
where $2 E(\om)= \hbar\om\coth(\hbar\om/2T)$ and
$D_{ijmn}=D_t(\delta_{im}\delta_{jn}+\delta_{in}\delta_{jm}) +
(D_l-2D_t)\delta_{ij}\delta_{mn}$. The mixed correlation vanishes:
$<\G\, s_{ij}>=0$. In \cite{LL6} the correlation functions are obtained using
their relation (e.g.~\S 121 of \cite{LL5}) to coefficients appearing in the
expression for the rate of change of the entropy (see \S119 of \cite{LL5}).

The appearance of the factor $E(\om)$ implies that in the quantum mechanical
case, i.e.~for $\hbar\ne 0$, the correlation function exhibits, in general, a
finite correlation time. Furthermore, $E(\om)$ reminds one that diffusion
constants $D_l,\, D_t$ as well as thermal conductivity $\kappa$ are functions
of frequency, the even parts of which decrease for large frequencies
asymptotically, at least, as $1/\om$ (for the right hand sides of (\ref{corrs})
and (\ref{corrg}) to be proper Fourier transforms). This feature ensures
convergence of frequency integrals which will have to be dealt with in the
following.

The structure of the equations for the averages follow by putting the right
hand side of (\ref{linhyd1})--(\ref{linhyd3}) equal to zero. The equations of
motion for the fluctuations of $\delta\rho$, $\g$ and $\delta q$ can be
obtained in different ways. One of them is to directly use the equations
(\ref{linhyd1})--(\ref{linhyd3}) for the dynamical variables exploiting the
relations (\ref{corrs}), (\ref{corrg}) for the fluctuating forces. This one
will be applied later. An other possibility is the one described in section 2.
For the latter one needs to know the corresponding response functions, as the
density density response.

If external fields are introduced in the equations of motion according to the
rules
\begin{eqnarray}\label{coupext}
P \rightarrow & P + \rho_0\, U^{\rm e}_\rho \nonumber\\*
\g \rightarrow & \g +\rho_0\, {\bf U}^{\rm e} \\*
\delta T \rightarrow & \delta T + T_0\, U^{\rm e}_q\nonumber
\end{eqnarray}
then the interaction energy takes the form \cite{KadMart}:
\begin{equation}\label{pertham}
H^{\rm ext}=\int d\r\left(\delta\rho\, U^{\rm e}_\rho + \g\, {\bf U}^{\rm e}
+ \delta q\, U^{\rm e}_q \right)
\end{equation}
This is exactly of form (\ref{hint}). Therefore, the equations for the
response functions can be obtained taking the functional derivatives of
(\ref{linhyd1})--(\ref{linhyd3}) with respect to the external fields introduced
according to (\ref{coupext}). For example, the set of equations containing the
density density response
$\chi_{\rho\rho}(\k,\kp,\om)=(2\pi)^3\delta(\k-\kp)\chi_{\rho\rho}(\k,\om)$
is found by calculating the derivative with respect to $U^{\rm e}_\rho$. The
solution reads:
\begin{equation}\label{resfurr}
\chi_{\rho\rho}(\k,\om)= \rho_0\, \frac{-k^2}
{\om^2+i\om D_l k^2 - c^2 k^2 + i d k^4 c^2 \frac{1-c_v/c_p}{\om +i d k^2}}
\end{equation}
if the following abbreviations are used for derivatives of thermodynamic
functions:
\begin{equation}\label{defs}
d=\frac{\kappa}{\rho c_v}\qquad
c_x = \frac{T}{V\rho}\left(\frac{\partial S}{\partial T}\right)_x\qquad
c^2 =\left(\frac{\partial P}{\partial \rho}\right)_S\qquad
a= \left(\frac{\partial T}{\partial \rho}\right)_S
\end{equation}

The other response functions are found to be related to $\chi_{\rho\rho}$ in
the following way:
\begin{equation}\label{resfugr}
\chi_{{\rm g}_i\rho}(\k,\om)=\chi_{\rho{\rm g}_i}(\k,\om)=
k_i\,\frac{\om}{k^2}\,\chi_{\rho\rho}(\k,\om)
\end{equation}
\begin{equation}\label{resfuqr}
\chi_{q\rho}(\k,\om)=\chi_{\rho q}(\k,\om)=
\frac{-i\kappa a k^2}{\om+i d k^2}\,\chi_{\rho\rho}(\k,\om)
\end{equation}
\begin{equation}\label{resfuqg}
\chi_{q{\rm g}_i}(\k,\om)= \chi_{{\rm g}_i q}(\k,\om) =
k_i\, \frac{\om}{k^2}\,\chi_{q\rho}(\k,\om)
\end{equation}
\begin{equation}\label{resfuqq}
\chi_{qq}(\k,\om)= \frac{-i \kappa a k^2}{\om+i d k^2}\,\chi_{q\rho}(\k,\om)+
\frac{i \kappa T_0 k^2}{\om + i d k^2}
\end{equation}
\begin{equation}\label{resfugg}
\chi_{{\rm g}_i{\rm g}_j}(\k,\om)= \frac{k_i k_j}{k^2}\, \chi_l(\k,\om) +
\left(\delta_{ij} -\frac{k_i k_j}{k^2}\right)\, \chi_t(\k,\om)
\end{equation}
\begin{equation}\label{resful}
\chi_l(\k,\om)= \frac{\om^2}{k^2}\, \chi_{\rho\rho}(\k,\om) +\rho_0
\end{equation}
\begin{equation}\label{resfut}
\chi_t(\k,\om)= \rho_0\, \frac{i D_t k^2}{\om +i D_t k^2}
\end{equation}
{}From these functions the equilibrium values of the corresponding fluctuations
can be calculated -- at least the contributions $\tilde\sigma^{\rm st}$
(\ref{sigtilst}) -- combinations of which form the diffusion coefficients.

According to (\ref{eomfluct}), with the operator $\cal D$ defined by
(\ref{eomav}) and (\ref{linhyd1})--(\ref{linhyd3}), the equations of motion for
the fluctuations read:
\begin{equation}\label{eomsigrr}
\ddt\sigma_{\rho\rho}(\k,\kp)+i\,k_j\,\sigma_{{\rm g}_j\rho}-
i\, k^\prime_j\,\sigma_{\rho{\rm g}_j} = 0
\end{equation}
\begin{equation}\label{eomsigrg}
\ddt\sigma_{\rho{\rm g}_i}(\k,\kp)+i\, k_j\,\sigma_{{\rm g}_j{\rm g}_i}-
i\, k^\prime_j \,\sigma_{\rho\Pi_{ij}} = 2\, d_{\rho {\rm g}_i}
\end{equation}
\begin{equation}\label{eomsigrq}
\ddt\sigma_{\rho q}(\k,\kp) + i\, k_j\,\sigma_{{\rm g}_j q} +
\kappa\,{k^\prime}^2\,\sigma_{\rho T} = 2\, d_{\rho q}
\end{equation}
\begin{equation}\label{eomsiggg}
\ddt\sigma_{{\rm g}_i{\rm g}_j}(\k,\kp) +i\, k_l\,\sigma_{\Pi_{il}{\rm g}_j}-
i\, k^\prime_l\,\sigma_{{\rm g}_i\Pi_{jl}}= 2\, d_{{\rm g}_i{\rm g}_j}
\end{equation}
\begin{equation}\label{eomsiggq}
\ddt\sigma_{{\rm g}_i q}(\k,\kp) +i\, k_j\,\sigma_{\Pi_{ij} q} +
\kappa\,{k^\prime}^2\,\sigma_{{\rm g}_i T} = 2\, d_{{\rm g}_i q}
\end{equation}
\begin{equation}\label{eomsigqq}
\ddt\sigma_{qq}(\k,\kp) + \kappa\, k^2\,\sigma_{T q}+
\kappa\,{k^\prime}^2\,\sigma_{q T} = 2\, d_{qq}
\end{equation}
To simplify notation the arguments $\k,\kp$ are written only in those terms
containing a derivative with respect to time. Please notice that in the
present case the relation (\ref{hermsig}) translates to
$\sigma_{AB}(\k,\kp)=\sigma_{BA}(-\kp,-\k)$, where $A,B=\rho,\g,q$. To
shorten the notation further the abbreviations
\begin{equation}\label{sigapi}
k_l\,\sigma_{\Pi_{il}A}(\k,\kp)= k_i\,\sigma_{P A}(\k,\kp) -
i D_{imjn} k_m k_n\, \sigma_{{\rm g}_j A}(\k,\kp)
\; ,
\end{equation}
\begin{equation}\label{sigap}
\sigma_{P A} = c^2\, \sigma_{\rho A}+ \frac{\rho_0}{T_0}\, a\, \sigma_{q A}
\end{equation}
and
\begin{equation}\label{sigat}
\sigma_{A T}=a\, \sigma_{A\rho}+\frac{1}{\rho_0 c_v}\, \sigma_{A q}
\end{equation}
are introduced for certain combinations of fluctuations.

The diffusion coefficients follow either from the general formula
(\ref{defdifco}) together with (\ref{eomav}), (\ref{linhyd1})--(\ref{linhyd3})
or from (\ref{eomsigrr})--(\ref{eomsigqq}):
\begin{equation}\label{drg}
2\, d_{\rho {\rm g}_i}(\k,\kp)=
i\, k_j\,\sigma^{\rm st}_{{\rm g}_j{\rm g}_i} -
i\, k^\prime_j \,\sigma^{\rm st}_{\rho\Pi_{ij}}
=i\, k_j\,\sigma^{\rm st}_{{\rm g}_j{\rm g}_i} -
i\, k^\prime_i \,\sigma^{\rm st}_{\rho P}
\end{equation}
\begin{equation}\label{drq}
2\, d_{\rho q}(\k,\kp) =
i\, k_j\,\sigma^{\rm st}_{{\rm g}_j q} +
\kappa\,{k^\prime}^2\,\sigma^{\rm st}_{\rho T}
= \kappa\,{k^\prime}^2\,\sigma^{\rm st}_{\rho T}
\qquad\qquad
\end{equation}
\begin{equation}\label{dgg}
2\, d_{{\rm g}_i{\rm g}_j}(\k,\kp)=
i\, k_l\,\sigma^{\rm st}_{\Pi_{il}{\rm g}_j} -
i\, k^\prime_l\,\sigma^{\rm st}_{{\rm g}_i\Pi_{jl}}
= D_{imln}k_m k_n\sigma^{\rm st}_{{\rm g}_l{\rm g}_j}+
D_{lmjn}k^\prime_m k^\prime_n\sigma^{\rm st}_{{\rm g}_i{\rm g}_l}
\end{equation}
\begin{equation}\label{gq}
2\, d_{{\rm g}_i q}(\k,\kp) =
i\, k_j\,\sigma^{\rm st}_{\Pi_{ij} q} +
\kappa\,{k^\prime}^2\,\sigma^{\rm st}_{{\rm g}_i T}
=i\, k_i\,\sigma^{\rm st}_{P q}
\qquad\qquad
\end{equation}
\begin{equation}\label{dqq}
2\, d_{qq}(\k,\kp) =
\kappa\,k^2\,\sigma^{\rm st}_{T q}+\kappa\,{k^\prime}^2\,\sigma^{\rm st}_{q T}
\qquad\qquad\qquad\qquad
\end{equation}
where $\sigma^{\rm st}_{AB}\equiv\sigma^{\rm st}_{AB}(\k,\kp)$.
To obtain these results use was made of the behaviour of $\rho, q, \g$
under time reversal.

As mentioned earlier in this section, knowing the equations
(\ref{linhyd1})--(\ref{linhyd3}) for the dynamical variables one can determine
the equations for the fluctuations and the diffusion coefficients
in a second way. Realizing that the time dependence of $\sigma_\nm(t)$ can
be interpreted as average of time dependent dynamical
variables $\hat A_\nu(t)$
%(in quantum mechanics one would speak of the Hei\ss enberg picture)
%\begin{equation}\label{dtsig}
%\sigma_\nm(t)=\left<\frac{1}{2}[\hat A_\nu(t)\, ,\,
%\hat A_\mu^\dagger (t)]_+\right> - A_\nu(t) A_\mu^\ast(t)
%\; ,
%\end{equation}
one can calculate the time derivatives of the fluctuations by inserting the
expressions for time derivatives of dynamical variables. The resulting
equations for the fluctuations are identical to
(\ref{eomsigrr})--(\ref{eomsigqq}). For the diffusion coefficients one
finds the following explicit formulas:
\begin{equation}\label{landrg}
2\, d^{\rm L}_{\rho {\rm g}_i}(\k,\kp)= -i\,k^\prime_j\,\frac{1}{2}
\left<\Big[\delta\hat\rho(\k,t)\, ,\, s_{ij}(-\kp,t)\Big]_+\right>
\end{equation}
\begin{equation}\label{landrq}
2\, d^{\rm L}_{\rho q}(\k,\kp)= i\, k^\prime_j\frac{1}{2}
\left<\Big[\delta\hat\rho(\k,t)\, ,\, G_j(-\kp,t)\Big]_+\right>
\end{equation}
\begin{equation}\label{landgg}
2\, d^{\rm L}_{{\rm g}_i{\rm g}_j}(\k,\kp)=
i\, k_l \frac{1}{2}\left<\Big[s_{il}(\k,t)\, ,\,
{\rm\hat g}_j(-\kp,t)\Big]_+\right> -i\, k^\prime_l \frac{1}{2}
\left<\Big[{\rm\hat g}_i(\k,t)\, ,\, s_{jl}(-\kp,t)\Big]_+\right>
\end{equation}
\begin{equation}\label{landgq}
2\, d^{\rm L}_{{\rm g}_i q}(\k,\kp)=
i\, k_j \frac{1}{2}\left<\Big[s_{ij}(\k,t)\, ,\, \hat q(-\kp,t)\Big]_+\right>+
i\, k^\prime_j\frac{1}{2}
\left<\Big[{\rm\hat g}_i(\k,t)\, ,\, G_j(-\kp,t)\Big]_+\right>
\end{equation}
\begin{equation}\label{landqq}
2\, d^{\rm L}_{qq}(\k,\kp)= -i\,k_j\,\frac{1}{2}
\left<\Big[G_j(\k,t)\, ,\, \hat q(-\kp,t)\Big]_+\right> + i\, k^\prime_j
\frac{1}{2}\left<\Big[\hat q(\k,t)\, ,\, G_j(-\kp,t)\Big]_+\right>
\end{equation}
To evaluate them one has to express $\delta\hat\rho, \hat\g, \delta\hat q$ as
functions of the fluctuating forces with the help of
(\ref{linhyd1})--(\ref{linhyd3}). Using the response functions (\ref{resfurr}),
(\ref{resfugr})--(\ref{resfut}) one can summarize the result by
\begin{equation}\label{solava}
\hat A(\k,\om)=
\chi_{A\rho}(\k,\om)\,\frac{k_i k_j}{\rho_0\, k^2}\, s_{ij}(\k,\om) +
\chi_{A q}(\k,\om)\,\frac{-i k_i}{\kappa T_0\, k^2}\, G_i(\k,\om)
\; ,
\end{equation}
where $A=\rho,\k\g, q$, and
\begin{equation}\label{solavt}
\Big(\g_t(\k,\om)\Big)_i=
\chi_t(\k,\om)\,\frac{i}{\rho_0\, D_t\, k^2}\Big(k_j\, s_{ij}(\k,\om)\Big)_t
\; .
\end{equation}
Using (\ref{solava}), (\ref{solavt}) as well as the correlation functions
(\ref{corrs}) and (\ref{corrg}) one finds that the diffusion coefficients
(\ref{landrg})--(\ref{landqq}) are diagonal in $\k,\kp$:
\begin{equation}\label{difpdiag}
d^{\rm L}_{AB}(\k,\kp)= (2\pi)^3\delta(\k-\kp)\, d_{AB}(\k)
\end{equation}
where $A,B=\rho,\g,q$. For the $d(\k)$ one ends up with the following
expressions:
\begin{eqnarray}\label{landexpgr}
d_{\rho{\rm g}_i}(\k) &=&
\int\frac{d\om}{2\pi} E(\om)\, (-i D_l)\,k_i\,\chi_{\rho\rho}(\k,\om)\\*
d_{\rho q}(\k) &=& \int\frac{d\om}{2\pi} E(\om)\chi_{\rho q}(\k,\om)\\*
d_{{\rm g}_i q}(\k) &=& \int \frac{d\om}{2\pi}
E(\om)\, \frac{\om+iD_l k^2}{\om}\,\chi_{{\rm g}_i q}(\k,\om)\\*
d_{qq}(\k) &=& \int \frac{d\om}{2\pi} 2 E(\om)\chi_{qq}(\k,\om)\\*
d_l(\k) &=&
\int\frac{d\om}{2\pi} 2 E(\om)(-i D_l)\, k_i\,\chi_{{\rm g}_i \rho}(\k,\om)\\*
d_t(\k) &=& \int \frac{d\om}{2\pi} 2 E(\om) \chi_t(\k,\om)
\end{eqnarray}
With the help of the expressions (\ref{resfurr}),
(\ref{resfugr})--(\ref{resfut}) for the response functions one finds the same
results for the diffusion coefficients (\ref{drg})--(\ref{dqq}), provided that
for $\sigma^{\rm st}$ only the contribution $\tilde\sigma^{\rm st}$ is
inserted, i.e.
\begin{equation}\label{difdiag}
\tilde d_{AB}(\k,\kp)= (2\pi)^3\delta(\k-\kp)\, d_{AB}(\k)=
d^{\rm L}_{AB}(\k,\kp)
\; ,
\end{equation}
with $A,B=\rho,\g,q$, where the tilde above $d$ reminds of the
restriction on $\tilde\sigma^{\rm st}$.

We thus see, that the theory of fluctuations of Landau and Lifschitz leads to
the same expressions for the diffusion coefficients as our method, provided the
additional contribution $\Delta\sigma$ to $\sigma^{\rm st}$ (\ref{fdtgen})
either vanishes exactly or is neglected.

\section{Comparison with Boltzmann-Langevin theory}

In this section connection will be established between the approach explained
in section 2 and the Boltzmann-Langevin theory by comparing the expressions for
the diffusion coefficients. To calculate the latter with the method of section
2, in general one has to know the behaviour of the response function as
function of frequency. However, investigating (\ref{sigtilst}) one finds that
less information is needed to exploit the fluctuation dissipation theorem in
the classical limit: one only has to determine the static response, i.e.~the
value of the response function at $\om=0$. From (\ref{eomav}) and
(\ref{chimakro}) the equation for the latter follows as:
\begin{equation}\label{eqstares}
{\cal D}_{\nu\nup}\chi_{\nup\mu}(\om =0)=
\frac{\delta c_\nu}{\delta U^{\rm ext}_\mu(\om=0)}
\end{equation}
The combination on the left hand side of (\ref{eqstares}) appears also in the
expression (\ref{defdifco}) for the diffusion coefficient since the
contribution $\tilde\sigma^{\rm st}$ (\ref{sigtilst}) to the latter equals, in
the classical limit, the product of static response and temperature. Therefore,
it is not necessary to solve (\ref{eqstares}) for the static response -- only
the right hand side of (\ref{eqstares}) is needed.

Let us now look at the cases of BUU and Landau equation as the ones for the
averages, i.e.~the ones defining $c_\nu$ via (\ref{eomav}).
Linearizing the equation for the distribution function in the deviation from
the homogeneous equilibrium distribution one finds
\begin{equation}\label{eomlin}
\ddt\,\delta n_\p(\r,t)+\left({\bf v}_\p\nabla_r -J_{\p,\r}\ast\right)
\left(\delta n_\p(\r,t) -\dern0\delta\varepsilon_\p(\r,t)\right)= 0
\; .
\end{equation}
The linear operator $J$ is defined by the collision term:
\begin{equation}\label{defig}
I[n_\p]\Big\vert_{\rm lin} =
J_{\p,\r}\ast\Big(n_\p(\r,t)-n^0(\varepsilon_\p)\Big)
\end{equation}
The asterix reminds that, in general, $J_{\p,\r}$ includes integrations over
momentum and position. (In the non-Markov case, there will be an additional
integration over time. An index $t$ at $J$, however, is omitted for the
moment.)

Please notice that the energy appearing in the argument of the equilibrium
distribution $n^0$ is taken to be the momentary one. The latter differs from
$\varepsilon_p^0$, which appears in (\ref{rela}) and which represents the
energy in equilibrium without external field, by the amount
\begin{equation}\label{delspen}
\delta\varepsilon_\p=\int\frac{d\pp}{h^3} f_{\p\pp}\delta n_\pp +U^{\rm ext}_\p
\; .
\end{equation}
The kernel $f_{\p\pp}$ is given by $f_{\p\pp}= \partial U/\partial \rho$ for
the case of the Boltzmann equation with a momentum independent mean field
$U[\rho]$. In Landau Fermi Liquid theory $f_{\p\pp}$ is called quasi particle
interaction.

To calculate the diffusion coefficient one has to determine $c_\nu$ as
functional of the external field. Comparing (\ref{eomlin}) with the general
form (\ref{eomav}) of the equation for the averages one obtains
\begin{equation}\label{relc}
c_\nu\quad \leftrightarrow\quad \Big({\bf v}_\p\nabla_r -J_{\p,\r}\ast\Big)
\left(-\dern0\,U^{\rm ext}_\p(\r,t)\right)
\; .
\end{equation}
The right hand side of the equation (\ref{eqstares}) for the static response
corresponds therefore to
\begin{equation}\label{relcabl}
\frac{\delta c_\nu}{\delta U^{\rm ext}_\mu(\om=0)} \quad\leftrightarrow\quad
\Big({\bf v}_{\p_\nu}\nabla_{r_\nu} - J_{\p_\nu,\r_\nu}\ast\Big)
\left(-\frac{\partial n^0}{\partial\varepsilon_{p_\nu}}\,
h^3\delta(\p_\nu-\p_\mu)\delta(\r_\nu-\r_\mu) \right)
\; .
\end{equation}
Denoting by $\tilde d$ the contribution from $\tilde\sigma^{\rm st}$
(\ref{sigtilst}) to the diffusion coefficient (\ref{defdifco}), one finds from
(\ref{relcabl})
\begin{equation}\label{dcl}
\left. 2\tilde d(\p,\r,\pp,\rp)\right\vert_{\rm cl} =
-T\left(J_{\p,\r}\ast\left(-\dern0 h^3\delta(\p-\pp)\delta(\r-\rp)\right)+
\Big(\r,\p\,\leftrightarrow\, \rp,\pp\Big) \right)
\end{equation}
since the terms with the spatial derivative in (\ref{relcabl}) cancel because
of $(\nabla_r+\nabla_{r^\prime})\delta(\r-\rp)=0$.

For a local, Markov collision term the linear operator $J$ (\ref{defig}) can be
written as the sum of a diagonal and a non-diagonal contribution
\begin{equation}\label{sepstos}
J_{\p,\r}\ast h(\p,\r,t) = -\frac{1}{\tau_\p}\, h(\p,\r,t) +
\int\frac{d\pp}{h^3} I_{\p\pp}\, h(\pp,\r,t)
\end{equation}
where $h(\p,\r,t)$ is some function of $\p,\r,t$. The action of the diagonal
part of $J$ consists of a multiplication with a function of $\p$ only which,
in (\ref{sepstos}), is written as $1/\tau_\p$ to indicate, that this factor has
the dimension of an inverse time. It is uniquely determined by $J$ under the
subsidiary condition that the kernel $I_{\p\pp}$ does not contain terms
proportional to $\delta(\p-\pp)$.

For the collision term (\ref{sepstos}) the diffusion coefficient follows as:
\begin{equation}\label{tildfin}
\left. \tilde d(\p_1,\r_1,\p_2,\r_2)\right\vert_{\rm cl} = \delta(\r_1-\r_2)
\left[\frac{n^0_{p_1}{\overline n}^0_{p_1}}{\tau_{\p_1}}\, h^3\delta(\p_1-\p_2)
-\frac{1}{2}\left(I_{\p_1\p_2}n^0_{p_2}{\overline n}^0_{p_2} +
I_{\p_2\p_1}n^0_{p_1}{\overline n}^0_{p_1}\right)\right]
\; ,
\end{equation}
using the definition ${\overline n}_\p=1-n_\p$.

This result can be compared with the one obtained in the Boltzmann-Langevin
theory for the (Markovian) BUU collision term.
The comparison will be performed in two steps: In a first step results obtained
in \cite{RanAy} for the low temperature limit are used. In a second step the
restriction to small temperatures will be removed.

In \cite{RanAy} the Boltzmann-Langevin expressions for collision term and
diffusion coefficient were evaluated in the low temperature limit and for small
deviations from thermal equilibrium. The result is written in formula (21) of
\cite{RanAy}. To perform the comparison one first has to determine $\tau_\p$
and $I_{\p\pp}$ as defined by (\ref{sepstos}). Comparing \ref{sepstos} with
the expression for the collision term in \cite{RanAy} (there called drift
coefficient) one finds that, in the approximation of \cite{RanAy}, $1/\tau_\p$
is given by the momentum independent quantity $W_0$ of \cite{RanAy} and the
kernel $I_{\p_1\p_2}$ is given by $f^0_1 {\overline f}^0_1 C_{12}$ of
\cite{RanAy}. With the help of these relations the expression (\ref{tildfin})
turns into the lower one in formula (21) of \cite{RanAy}, i.e.~into the
Boltzmann-Langevin result for the diffusion coefficient in equilibrium and in
the low temperature limit.

However, the restriction to low temperature can be released. To this end one
starts from the general Boltzmann-Langevin expressions for drift- and diffusion
coefficient \cite{BCR,AG}. One finds that the contribution to the collision
term which is linear in $\delta n_\p$ can be written as (see appendix B)
\begin{equation}\label{blcollterm}
I[n_\p]\vert_{\rm lin}=
-{\alpha^2[n^0](\p)\over 2 n^0_p {\overline n}^0_p} \delta n_\p -
\int {d\p_2\over h^3}
{\alpha_{\rm cov}[n^0](\p,\p_2)\over 2 n^0_{p_2}{\overline n}^0_{p_2}}
\delta n_{\p_2}
\end{equation}
where (in the notation of \cite{BCR})
$\alpha[n](\p,\r,\pp,\rp)=\delta(\r-\rp)\Big(\alpha^2[n](\p)\, h^3
\delta(\p-\pp)+ \alpha_{\rm cov}[n](\p,\pp)\Big)$ appears as inhomogeneity in
the equation for the fluctuations, i.e.~is the equivalent to $2 d(\p,\pp)$ of
the present paper. Comparing (\ref{blcollterm}) with (\ref{sepstos}) one finds
the identification:
\begin{equation}\label{ident}
\left.
\begin{array}{l} \rm Boltzmann-\\ \rm Langevin\end{array}\qquad
\begin{array}{c} \alpha^2[n^0](p) \\ \\ \alpha_{\rm cov}[n^0](\p_1,\p_2)
%f^0_1 {\overline f}^0_1 C_{12}
\end{array}\quad
\right\}\quad\leftrightarrow\quad\left\{\quad
\begin{array}{c}2 n^0_p {\overline n}^0_p\, /\,\tau_\p \\ \\
-2\, I_{\p_1\p_2} n^0_{p_2}{\overline n}^0_{p_2} \end{array}\right.\qquad
\begin{array}{l} \rm present \\ \rm work \end{array}
\end{equation}
Inserting (\ref{ident}) in the expression (\ref{tildfin}) results in
\begin{equation}\label{relbldcl}
2 \tilde d(\p,\r,\pp,\rp)\vert_{\rm cl}= \alpha[n^0](\p,\r,\pp,\rp)
\end{equation}
one only has to use the symmetry property
$\alpha_{\rm cov}[n^0](\p,\pp)=\alpha_{\rm cov}[n^0](\pp,\p)$.
Thus, the equilibrium diffusion coefficient of the Boltzmann-Langevin theory
turns out identical to the contribution $\tilde d$ to the one of the present
approach provided the latter is calculated with the classical form of the
fluctuation dissipation theorem, i.e.~is identical to (\ref{tildfin}).

As mentioned before, in the Boltzmann-Langevin theory the correlation function
for the fluctuating force, $C(\p,\r,t,\pp,\rp,t^\prime)$, when calculated in
equilibrium  and for a Markovian collision term, is given \cite{AG,RR} by
\begin{equation}\label{blff}
C(\p,\r,t,\pp,\rp,t^\prime)=
\left. 2\,\tilde d(\p,\r,\pp,\rp)\right\vert_{\rm cl} \delta(t-t^\prime)
\; ,
\end{equation}
where the relation between the Boltzmann-Langevin diffusion coefficient and
(\ref{tildfin}) was utilized. Using the method of Landau and Lifschitz,
applied in the previous section, to calculate the correlation function of the
fluctuating force for BUU or Landau equation (for the latter see
e.g.~\cite{AbKh}) one obtains a result that agrees with (\ref{blff}), again
for the classical fluctuation dissipation theorem and for a Markovian collision
term.

In a recent work \cite{Ayik} the Boltzmann-Langevin theory is extended to a
non-Markovian collision term. In the framework of the theory of Landau and
Lifschitz this generalization can be done without difficulty, since until the
final step in the derivation one does not need to specify the form of the
collision term. Indeed, one finds for $C(\p,\r,t,\pp,\rp,t^\prime)$ an
expression similar to the right hand side of (\ref{dcl}). However, in the
former an $\delta$-function $\delta(t-t^\prime)$ appears besides those in the
positions and in the momenta, and the operator $J$ (\ref{defig}) is now non
local in time:
\begin{equation}\label{ignonloc}
J_{\p,\r,t}\ast h(\p,\r,t) =
\int\frac{d\pp d\rp dt^\prime}{h^3}\,
{\cal I}(\p,\r,t;\pp,\rp,t^\prime)\, h(\pp,\rp,t^\prime)
\end{equation}
Following \cite{LL6} the quantum correlation function of the fluctuating force
is obtained multiplying the Fourier transform (in time) of the classical one
by $E(\om)/T$. Thus one finds the former to read:
\arraycolsep1.5pt
\begin{eqnarray}\label{corrfkt}
C(\p,\r,t,\pp,\rp,t^\prime)&=&\int\frac{d\om}{2\pi}e^{-i\om t}
E(\om)\int d\tau e^{i\om\tau} J_{\p,\r,\tau}\ast
\left(\dern0 h^3\delta(\p-\pp)\delta(\r-\rp)\delta(\tau-t^\prime)\right)+
\nonumber\\*
&&+\, \Big(\p,\r,t \leftrightarrow \pp,\rp,t^\prime\Big)
\end{eqnarray}
In the expression for $C$ the use of the kernel $\cal I$ (\ref{ignonloc}) is
omitted for the sake of an easier comparison with (\ref{dcl}) in the Markov
limit.

It is straightforward to show that (\ref{corrfkt}) leads to the result found
in \cite{Ayik} for the correlation function of the Boltzmann-Langevin theory
in equilibrium, provided the expression for the non Markovian collision term
of \cite{Ayik} is used to define $J$ according to (\ref{defig}).

\section{Sum rules -- the case of zero wave vector}

In finite, closed systems the total number of particles, the total momentum and
the total energy are conserved quantities. Starting with an ensemble of systems
all having the same particle number, momentum and energy, the dynamical
variables corresponding to the latter should not exhibit any fluctuations at a
later time. This feature constrains the possible forms for the equations of
motion of the fluctuations. Especially, it implies sum rules for fluctuations
$\sigma$ and diffusion coefficient $d$. Since the deviations of total particle
number, momentum or energy from the corresponding equilibrium values are given
by the values of the phase space integral of the deviation of the distribution
function weighted with  $1$, $\p$ or $\eps_\p$, respectively, the sum rules
read:
\begin{equation}\label{sumrules}
\int\frac{d\r d\p}{h^3}\, O_\p\, b(\p,\r,\p,\rp)= 0
\end{equation}
where $O_\p=1,\p,\eps_\p$ and $b=\sigma, d$. In the following it will be
demonstrated that the results for $\tilde\sigma$ and $\tilde d$ obtained with
the method of section 2 together with the assumption $\Delta\sigma=0$ are
consistent with these sum rules.

At first, the classical limit shall be addressed. Here, the diffusion
coefficient obtained in section 4 was found to agree with the result of the
Boltzmann-Langevin theory in equilibrium. The former is, therefore, known
\cite{AG,BCR} to fulfil the sum rules (\ref{sumrules}).

For investigation of the fluctuations one might think of solving
(\ref{eqstares}) for the static response. However, there is some subtlety
involved: response functions may be non analytic at $\om =0=\k$ and one has
to choose carefully the correct sequence in performing limits. This fact is
relevant for the present case since in the sum rules (\ref{sumrules}) there
occurs an integral over space. The value of the latter is thus equal to the one
of the Fourier transform of the integrand taken at $k=0$. Moreover, in the
contribution $\tilde\sigma^{\rm st}$ to the integrand there appears, in the
classical limit, the response function at $\om=0$. The order in which the two
limits -- $k,\om\to 0$ -- have to be taken can be found as follows. As said
before, the sum rules arise due to the conservation laws for particle number,
momentum and energy. The deviations of the latter from their equilibrium values
are the averages of dynamical variables given by
\begin{equation}\label{dynvarint}
\hat M_{O_p} (k=0) = \int\frac{d\r d\p}{h^3}\, O_\p\, \delta\hat n_\p(\r)
\quad ,
\end{equation}
where $O_\p=1,\p,\eps_\p$ respectively and
$<\delta\hat n_\p(\r)>=\delta n_\p(\r)$.
Due to the conservation laws $\hat M_{O_p}(k=0)$ does not exhibit any
fluctuations:
\begin{equation}\label{nofluc}
\sigma_{M_{O_p}(k=0), A} = 0
\quad ,
\end{equation}
where $\hat A$ is some dynamical variable. Inserting here (\ref{dynvarint})
one recovers (\ref{sumrules}) for $b=\sigma$ when $A$ is taken as $n_\p$.
The fluctuations are related to the static response whose microscopic form
is given by \cite{KadMart}
\begin{equation}\label{mresfu}
\chi_{M_{O_p}(k=0), A}(\om=0)=\int dt\, {i\over\hbar}
\left<[\hat M_{O_p}(k=0)\, ,\, A^\dagger(-t)]\right> \Theta(t)
\quad .
\end{equation}
The expression on the right hand side can be rewritten to give
\begin{equation}\label{mresfu2}
\chi_{M_{O_p}(k=0), A}(\om=0)=\lim_{\om\to 0}\int dt\, e^{i\om t}{i\over\hbar}
\left<\left[\lim_{k\to 0} \int\frac{d\r d\p}{h^3} e^{-i\k\r} O_\p
\delta\hat n_\p(\r)\, ,\, A^\dagger(-t)\right]\right> \Theta(t)
\quad .
\end{equation}
If the limit $k\to 0$ can be commuted both with the averaging procedure and
with the time integration, then one finds:
\begin{equation}\label{mresfu3}
\chi_{M_{O_p}(k=0), \pp}(\om=0)=\lim_{\om\to 0}\left(\lim_{k\to 0}
\int\frac{d\p}{h^3}\, O_\p\chi_{\p\pp}(\k,\om)\right)
\end{equation}
Therefore, the sum rules for the fluctuations ($b=\sigma^{\rm st}$) read in
the classical limit:
\begin{equation}\label{surul}
\Delta\sigma_{M_{O_p}(k=0),\pp} + T \lim_{\om\to 0}\left(\lim_{k\to 0}
\int\frac{d\p}{h^3}\, O_\p\chi_{\p\pp}(\k,\om)\right) = 0
\end{equation}
Thus, for calculating the left hand side of (\ref{sumrules}) one has first
to perform the momentum integral of the response function at finite $k,\om$
weighted with $O_\p$. Then one has to take the limit $k\to 0$ (corresponding
to the spatial integral) and, finally, $\om\to 0$. Due to this sequence of
limits the investigation about the fluctuations fulfilling the sum rules
turns out to be not easier in the classical limit than in the general case.

There is a further, indirect argument that the sum rules are related to the
order of limits stated, i.e.~first $k\to 0$ and then $\om\to 0$. It is well
known \cite{Forster}, that taking first the limit for $\om$ and then for $\k$
leads to thermodynamic derivatives, e.g.~to the compressibility in the case of
the density density response function. However, a finite, closed system can
have a finite compressibility, too. And indeed, solving (\ref{eqstares}) in
the case of Landau Fermi Liquid theory one obtains \cite{HKT} the known, finite
value for the compressibility \cite{BP}.

In the general (i.e.~not classical) case one may proceed as follows in order to
prove the sum rules to hold true for fluctuations and diffusion coefficient.
Let again $O_\p$ be either $1$, $\p$ or $\eps_\p$. Multiplying the equation of
motion for the distribution function $n_\p(\r,t)$ by $O_\p$ and integrating
with respect to $\r,\p$ one finds \cite{BP}
\begin{equation}\label{inteomlin}
\int\frac{d\r d\p}{h^3}\, O_\p\, \ddt \delta n_\p(\r)= 0
\; .
\end{equation}
To obtain this result one only has to integrate by parts the terms containing
the derivatives with respect to $\r$ and $\p$ and to use the feature of the
collision term that number of particles, momentum and energy are preserved.

Taking the functional derivative of (\ref{inteomlin}) with respect to the
external field results in
\begin{equation}\label{dtchiko}
\ddt \int\frac{d\r d\p}{h^3}\, O_\p\, \chi_{\p\pp}(\r,t,\rp,t^\prime)= 0
\; .
\end{equation}
Since the response function is proportional to the Heavyside-$\Theta$-function,
$\chi_{\p\pp}(\r,t,\rp,t^\prime)\sim \Theta(t-t^\prime)$ \cite{KadMart},
reflecting causality, the solution of (\ref{dtchiko}) reads
\begin{equation}\label{chiko}
\int\frac{d\r d\p}{h^3}\, O_\p\, \chi_{\p\pp}(\r,t,\rp,t^\prime)= 0
\; .
\end{equation}
Due to (\ref{defchippz}) the dissipative part of $\chi$ has the same property.
Therefore, it follows from (\ref{sigtilst}) that
\begin{equation}\label{sigko}
\int\frac{d\r d\p}{h^3}\, O_\p\, \tilde\sigma^{\rm st}_{\p\pp}(\r,\rp)=
\int\frac{d\p}{h^3}\, O_\p\, \tilde\sigma^{\rm st}_{\p\pp}(\k=0)= 0
\; .
\end{equation}
We thus see that the contribution $\tilde\sigma^{\rm st}$, when being
calculated from the fluctuation dissipation theorem, fulfills the same sum
rules as the fluctuations themselves. An approximation like $\Delta\sigma =0$,
in other words approximating $\sigma^{\rm st}$ by $\tilde\sigma^{\rm st}$, is,
therefore, consistent with the sum rules (\ref{sumrules}) for the fluctuations.

Let us now turn to the diffusion coefficient. According to (\ref{defdifco})
the latter is given by
\arraycolsep1.5pt
\begin{eqnarray}\label{difcolan}
d(\r,\p,\rp,\pp)&=& \Big({\bf v}_\p\nabla_r - J_{\p,\r}\ast\Big)
\left(\sigma^{\rm st}_{\p\pp}(\r,\rp)-\dern0 f_{\p\ppp}(\r,\rpp)
\sigma^{\rm st}_{\ppp\pp}(\rpp,\rp)\right)+ \nonumber\\*
&& +\, \Big(\,\p,\r\,\leftrightarrow\, \pp,\rp\,\Big)^\ast
\end{eqnarray}
where integration over double primed quantities is to be understood. The last
term on the right hand side stands for the complex conjugate of the first term
but with the unprimed variables exchanged with the single primed ones.
Multiplying $d$ with $O_\p$ and integrating the product with respect to
position and momentum one finds that the contribution from the first term on
the right hand side of (\ref{difcolan}) vanishes. This is due to the same
reasons that lead to (\ref{inteomlin}), i.e.~directly due to the structure of
the equation of motion for the distribution function. The other contribution
vanishes due to (\ref{sigko}) with (\ref{hermsig}), i.e.~due to features of the
fluctuations. Thus the sum rules (\ref{sumrules}) for $b=d$
\begin{equation}\label{intd}
\int\frac{d\r d\p}{h^3}\, O_\p\, d(\p,\r,\p,\rp)=
\int\frac{d\p}{h^3}\, O_\p\, d(\p,\pp,\k=0)= 0
\; .
\end{equation}
are proven.

The diffusion coefficient forms the inhomogeneity in the equation for the
(non equilibrium) fluctuations. Due to (\ref{intd}) the equation of motion for
the product of the fluctuations with $O_\p$ integrated over $\r,\p$ is
homogeneous. Therefore, the dynamic fluctuations obey the same sum rules as
$\tilde\sigma^{\rm st}$, i.e.~(\ref{sigko}), for all times provided
they did in the initial state.

It should be mentioned that these properties, especially the ones of the
fluctuations, are restricted to $\k=0$. Indeed, it turns out \cite{thesis},
that the limit $k\to 0$ of $\tilde\sigma^{\rm st}_{\p\pp}(\k)$ differs from
the value at $\k=0$. This means, that the fluctuations are not continuous
functions at vanishing wave vector.

%\pagebreak[6]
\section{Summary}

The problem to determine equations of motion of fluctuations and diffusion
coefficients was addressed. A method \cite{HKT,KH} for its solution using
knowledge only of the average dynamics was reviewed using a more general
formulation. The basic ingredient of the method is the fluctuation dissipation
theorem and a suitable extension of the latter to instabilities. This extension
consists in the prescription to determine the diffusion coefficients as the
analytic continuation of the expression valid for the stable system.

The connection between the diffusion coefficient in the present approach and
the correlation function of the fluctuating force in the one of Bixon and
Zwanzig \cite{BZ} was explained. The method was then applied to hydrodynamics.
Equation of motion of the fluctuations and diffusion coefficients were
demonstrated to be the same as those obtained using the theory of fluctuations
by Landau and Lifschitz \cite{LL6}. Subsequently, the case of BUU and Landau
equation as equation for the average one (quasi) particle distribution was
addressed. An expression for the diffusion coefficient in terms of the
linearized, Markov collision operator was calculated using the classical form
of the fluctuation dissipation theorem. This expression was found to coincide,
for BUU, with the one obtained for the equilibrium distribution in the
Boltzmann-Langevin theory \cite{RanAy}. The connection with both the theory by
Landau--Lifschitz and the Boltzmann-Langevin theory holds under the assumption
that for the equilibrium fluctuations -- and, consequently, their analytical
continuation -- only that contribution is taken into account which arises from
the frequency integral of the dissipative part of the response function
(weighted with $\hbar\coth(\hbar\om/2T)/\pi$).

In the framework of the theory of Landau and Lifschitz the generalization to
a non-Markovian collision term in Boltzmann and Landau equation was presented.
The correlation function of the fluctuating force obtained for the case of the
Boltzmann equation coincides with the equilibrium one of the extension
\cite{Ayik} of the Boltzmann-Langevin theory provided the same form is used for
the collision term.

To establish sum rules arising from conservation of particle number and
momentum the behaviour of the response functions for vanishing wave vector and
frequency was taken into account properly. Clarifying the sequence in which
the limits $k\to 0$ and $\omega\to 0$ have to be calculated, results of
previous work \cite{HKT} could be interpreted correctly.
\vspace{1cm}

\noindent
{\bf Acknowledgments}
\vspace{0.2cm}

\noindent
The author wants to thank H.~Hofmann for many discussions during the past
years, for helpful criticism and suggestions as well as for carefully reading
the manuscript. This work was supported in part by the Deutsche
Forschungsgemeinschaft as well as by the National Science Foundation under the
Grant No.~PY-9403666.

\vspace{1cm}

\newpage
\begin{appendix}
\setcounter{equation}{0}
\renewcommand{\theequation}{\mbox{\Alph{section}.\arabic{equation}}}
\noindent
{\Large \bf Appendix:}
\section{Detailed comparison with Bixon and Zwanzig}
In this appendix the comparison of the present approach with the one of Bixon
and Zwanzig \cite{BZ} summarized at the end of section 2 will be worked out in
more detail.

Bixon and Zwanzig start with the classical Boltzmann equation without mean
and/or external field as the equation for the average one particle distribution
function. Augmenting the linearized Boltzmann equation by a fluctuating force,
$f_B\, F(t)$ in their notation, the authors write the result in the form
\begin{equation}\label{linbolfluc}
{\partial \phi\over\partial t} + L \phi = F(t)
\end{equation}
where $\phi(t)$ is the relative deviation of the distribution from the
Boltzmann distribution. The correlation function of the fluctuating force is
written as
\begin{equation}\label{bzcorrfluc}
\left<F(\k_1,\v_1;t_1)F(\k_2,\v_2;t_2)\right>= 2 B(\k_1,\v_1,\k_2,\v_2)
\delta(t_1-t_2)
\; .
\end{equation}
For $B$ the authors find the expression
\begin{equation}\label{strengthfluc}
2 B(\k_1,\v_1,\k_2,\v_2)= (L_1+L_2)\left<\phi_1\phi_2\right>
\; ,
\end{equation}
where the average on the right hand side has to be performed in equilibrium.
For the calculation of the correlation function (\ref{bzcorrfluc}) the authors
use the expression
\begin{equation}\label{bzeqfluc}
\left<\phi(\k_1,\v_1)\phi(\k_2,\v_2)\right>= [f_B(\v_2)]^{-1}\delta(\v_1-\v_2)
\delta(\k_1+\k_2)
\end{equation}

After this repetition the comparison can be performed easily. This the more, as
the equation for the average distribution following from (\ref{linbolfluc}) by
averaging of both sides, is written directly in the form (\ref{eomav}): one
only has to identify $\hat A_\nu$ with $\phi(\k_\nu,v_\nu)$. Clearly, one has
\begin{equation}\label{reldl}
{\cal D}_\nm A_\mu \leftrightarrow (L\phi)(\k_\nu,\v_\nu)
\, .
\end{equation}
{}From (\ref{defdifco}), (\ref{strengthfluc}) and (\ref{reldl}) follows
\begin{equation}\label{reldifb}
d_\nm= B(\k_\nu,\v_\nu, -\k_\mu, \v_\mu)
\; ,
\end{equation}
where the minus sign in front of $k_\mu$ is due to the hermitian conjugation of
the second dynamical variable in the definition (\ref{deffluct}) of the
fluctuations. Thus the statement at the end of section 2 about the relation of
the diffusion coefficient and the correlation function of the fluctuating force
is proven.

In the rest of this appendix it will be demonstrated how the expression
(\ref{bzeqfluc}) for the equilibrium fluctuations in a classical ideal gas
can be obtained from the related response function. First it should be noted,
that in (\ref{bzeqfluc}) there appears the fluctuations at finite wave-vectors
$\k_1,\k_2$. Next one realizes that in the classical limit only the static
response is needed to obtain $\tilde\sigma$. From (\ref{eomlin}) one finds
after Fourier transformation of the time and putting $\om=0$:
\begin{equation}\label{omodistr}
\left(\v_\r\nabla_r-J_{\p,\r}\ast\right)
\left(\delta n_\p(\r,\om=0) -\dern0\delta\varepsilon_\p(\r,\om=0)\right)= 0
\end{equation}
The solution of this equation reads
\begin{equation}\label{soldistr}
\delta n_\p(\r,\om=0) -\dern0\delta\varepsilon_\p(\r,\om=0)= c_\p(\r)
\end{equation}
where $(\v_\r\nabla_r-J_{\p,\r}\ast)c_\p(\r)=0$. This seems to be no
progress at first site, but it is: $J$ does not depend on the external field;
consequently the same is valid for $c_\p(\r)$. Therefore, taking the functional
derivative of (\ref{soldistr}) with respect to $U^{\rm ext}(\rp,\om=0)$ one
finds the static response of non interacting particles after Fourier
transforming the spatial dependencies:
\begin{equation}\label{statresni}
\chi_{\p\pp}(\k,\kp,\om=0)= -\dern0 h^3 \delta(\p-\pp)(2\pi)^3\delta(\k-\kp)
\end{equation}
Since the derivative of the Boltzmann distribution with respect to the energy
equals the negative of the distribution divided by the temperature one ends up
with the result for the equilibrium fluctuations:
\begin{equation}\label{cleqflub}
\tilde\sigma(\p,\k,\pp,\kp)\vert_{\rm cl.~id.~gas}= n^0(p)h^3\delta(\p-\pp)
(2\pi)^3\delta(\k-\kp)
\end{equation}
where $n^0(p)$ is the equilibrium distribution function for the momentum with
the property $\int d^3p/h^3 n^0(p)= \int d^3 v f_B(v)$.
Therefore, $f_B(v)=(m/h)^3 n^0(m v)$ and from (\ref{cleqflub})
\begin{equation}\label{eqflucv}
\left<\delta f(\v_1,\k_1)\delta f(\v_2,\k_2)\right> = f_B(v_1)\delta(\v_1-\v_2)
(2\pi)^3\delta(\k_1+\k_2)
\end{equation}
Dividing both sides by $f_B(v_1)f_B(v_2)$ one finds (\ref{bzeqfluc}), beside a
factor $(2\pi)^3$ which is due to the different conventions of the Fourier
transform.
\newpage
\section{Collision term in BL-theory}
In this appendix the derivation of (\ref{blcollterm}) will be presented.
The notation used is the one of \cite{BCR}.

In a first step one observes that the linearized BUU collision term can be
written in the form:
\begin{equation}\label{applincol}
I[n_\p]\vert_{\rm lin} = \int (\prod_{i=2}^4 dp_i) {w_{12,34}\over 2}
\Big(\delta(\p-\p_3)-\delta(\p-\p_1)\Big) \Big[
(\delta n_1 n^0_2 +\delta n_2 n^0_1) {\overline n}^0_3 {\overline n}^0_4 -
(\delta n_3 {\overline n}^0_4+\delta n_4 {\overline n}^0_3)n^0_1 n^0_2 \Big]
\; ,
\end{equation}
where as short hand notation a number $i$ stands for $\p_i$ and, as before,
${\overline n}_p = 1- n_p$. Using the quantities $W^x$ introduced in \cite{BCR}
one finds:
\begin{eqnarray}\label{blformlincol}
I[n_\p]\vert_{\rm lin}=&&-\Big(W^+[n^0_p]+W^-[n^0_p]\Big)\delta n_p\nonumber\\
&&-\int d\p_2 \Big[{\overline n}^0_p\Big(W^{++}_0(\p,2)-W^{+-}_0(\p,2)\Big)+
            n^0_p\Big(W^{--}_0(\p,2)-W^{-+}_0(\p,2)\Big)\Big]\,\delta n_{\p_2}
\; ,\quad\nonumber \\
\end{eqnarray}
where the index $0$ indicates, that the corresponding quantity is to be taken
at $n^0$. Making use of the properties of the Fermi function $n^0$ as well as
of the energy conservation contained in the transition rate $w$ one can rewrite
the integrand in (\ref{blformlincol}) as
\begin{equation}\label{reltoz}
{\overline n}^0_p (W^{++}_0(\p,2)-W^{+-}_0(\p,2))+ n^0_p (W^{--}_0(\p,2)-
W^{-+}_0(\p,2))= {n^0_p Z^-_0(\p,2)+{\overline n}^0_p Z^+_0(\p,2)\over
2 n^0_2 {\overline n}^0_2}
\end{equation}
Again the definitions of the $Z^\pm$ are to be found in \cite{BCR}.

Remembering that the non diagonal part to $\alpha[n](\p,\pp)$ is related to the
$Z^\pm$ via
\begin{equation}\label{dcovbl}
\alpha_{\rm cov}[n^0](\p,\pp)=
n^0_p Z^-_0(\p,\pp)+{\overline n}^0_p Z^+_0(\p,\pp)
\end{equation}
while the diagonal part is given by
\begin{equation}\label{dvarbl}
\alpha^2[n^0](p) = 2 n^0_p {\overline n}^0_p\Big(W^+[n^0_p]+W^-[n^0_p]\Big)
\end{equation}
one obtains (\ref{blcollterm}) from (\ref{blformlincol}).

\end{appendix}

\end{document}